\documentclass[aip,jcp,reprint,amsmath,showkeys]{revtex4-1}
\usepackage{graphicx,dcolumn,bm,color,microtype,hyperref}

\newcommand{\br}{\bm{r}}

\newcommand{\mD}{\mathcal{D}}
\newcommand{\p}{\partial}
\newcommand{\alert}[1]{\textcolor{black}{#1}}

\DeclareMathOperator{\E}{E}
\DeclareMathOperator{\K}{K}

\begin{document}

\title{Three-electron coalescence points in two and three dimensions}

\author{Pierre-Fran{\c c}ois Loos}
\email{pf.loos@anu.edu.au}
\thanks{Corresponding author}
\author{Nathaniel J. Bloomfield}
\author{Peter M. W. Gill}	
\affiliation{Research School of Chemistry, Australian National University, Canberra ACT 2601, Australia}

\begin{abstract}
The form of the wave function at three-electron coalescence points is examined for several spin states using an alternative method to the usual Fock expansion.
We find that, in two- and three-dimensional systems, the non-analytical nature of the wave function is characterized by the appearance of logarithmic terms, reminiscent of those that appear as both electrons approach the nucleus of the helium atom.  The explicit form of these singularities is given in terms of the interelectronic distances for a doublet and two quartet states of three electrons in a harmonic well.
\end{abstract}


\maketitle

%----------------------------------------------------------------
\section{Introduction}
%----------------------------------------------------------------
Universal features of the electronic wave function $\Phi$ are of continued interest to physicists and chemists, as they guide the construction of highly accurate wave functions, \cite{Frankowski84, Freund84} explicitly correlated ans\"atze within F12 theory \cite{Hattig12, Kong12} and accurate Jastrow factors for quantum Monte Carlo (QMC) calculations. \cite{Huang97, Drummond04, LopezRios12}
The Coulombic singularity at short interelectronic distances $r_{12} = | \br_1 - \br_2 |$ dominates all other terms and, near the two-particle coalescence (2PC) point $r_{12} = 0$, the behavior of $\Phi$ becomes independent of other details of the system.

Early work by Kato, \cite{Kato51, Kato57} and elaborations by Pack and Byers-Brown, \cite{Pack66} showed that, as two opposite-spin electrons approach, $\Phi$ has the form
\begin{equation} \label{eq:kato}
	\Phi =  \left( 1 + \frac{r_{12}}{2} \right) \Phi(r_{12} = 0) + O(r_{12}^2).
\end{equation}
Similar 2PC conditions are known for triplet and unnatural parity singlet states. \cite{Pack66, Morgan93}

To remove divergences in the local energy $\Phi^{-1}\Hat{H}\Phi$ at the 2PC points, cusp conditions such as \eqref{eq:kato} must be satisfied.
These divergences are especially harmful in diffusion QMC calculations, where they can lead to population-control problems and significant biases. \cite{Drummond04}
Recently, several authors have further probed the coalescence behavior of the wave function, extending the analysis of Pack and Byers Brown to higher-order 2PC conditions. \cite{Rassolov96, Tew08, Kurokawa13, Kurokawa14}
Surprisingly, imposing these additional conditions in QMC calculations was not found to be beneficial. \cite{Drummond04, LopezRios12}

A wave function with correct 2PC behavior nonetheless yields a discontinuous local energy at three-particle coalescences (3PC), i.e.~where $r_{12} = r_{13} = r_{23} = 0$.  Unfortunately, a theoretical understanding of 3PC behavior has been much more elusive. \cite{Morgan86, HoffmannOstenhof92, Fournais05}
In 1954, Fock \cite{Fock54} rediscovered a result derived earlier by Gronwall \cite{Gronwall37} and Bartlett. \cite{Bartlett37}
Using hyperspherical coordinates, he showed that, when both electrons approach the nucleus in a helium atom, the wave function takes the form
\begin{equation} \label{eq:fock}
	\Phi = 1 - 2 (r_1+r_2) + \frac{r_{12}}{2} - \frac{\pi-2}{3\pi} \br_1 \cdot \br_2 \ln\left( r_1^2 + r_2^2 \right) + \ldots.
\end{equation}
Such logarithmic terms are characteristic of 3PC behavior \cite{Abbott87, Gottschalk87a, Gottschalk87b} and using a wave function based on \eqref{eq:fock} yields a continuous local energy for the helium atom. \cite{Myers91}

The present work aims to elucidate the form of the wave function at three-electron coalescences for three electrons, in various spin states, confined to a harmonic potential in $\mD = 2$ or $\mD = 3$ dimensions.
Eschewing the usual hyperspherical approach, we develop a new technique to show that, in each case, the wave function has non-analytical behavior characterized by a logarithmic term.
Atomic units are used throughout.

%----------------------------------------------------------------
\section{Theory}
%----------------------------------------------------------------
A system of three electrons, interacting coulombically but confined within a harmonic well, is variously called hookium, harmonium or Hooke's law atom.  It has been studied at high and low densities by Taut \textit{et al.} \cite{Taut03} and Cioslowski \textit{et al.} \cite{Cioslowski06, Cioslowski11, Cioslowski12}
The previous work of White and Stillinger, \cite{White71} and other work showing that the Kato cusp conditions are independent of the external potential, \cite{Kestner62, Taut93, Cioslowski00, TEOAS09, QuasiExact09, ExSpherium10, QR12} imply that our 3PC results will also hold for any external potential.

The Hamiltonian of the system is
\begin{equation}
	\Hat{H} = \frac{1}{2} \sum_{i=1}^3 \left( -\nabla^2_i + \omega^2\,r_i^2  \right) + \kappa \sum_{i<j}^3 \frac{1}{r_{ij}},
\end{equation}
where $\nabla_i^2$ is the $\mD$-dimensional Laplacian for electron $i$ and $r_{ij} = \left| \br_i - \br_j \right|$ is the interelectronic distance between electrons $i$ and $j$.
The strength of the interelectronic repulsion is measured by $\kappa$.
Without loss of generality, we choose the harmonic force constant $\omega^2 = 1$.

We consider only $S$ states (i.e.~$L = 0$) and, after adopting Jacobi coordinates, the center-of-mass coordinate
\begin{equation}
	\bm{\sigma} = (\br_1+\br_2+\br_3) / \sqrt{3}
\end{equation}
separates and one can write
\begin{equation} \label{eq:psi}
	\Psi(\bm{\sigma},\bm{\rho},\bm{\lambda}) = \psi(\rho,\lambda,\bm{\rho}\cdot\bm{\lambda})\Omega(\bm{\sigma}),
\end{equation}
where 
\begin{align}
	\bm{\rho} & = (\br_1 - \br_2) / \sqrt{2}	,
	&
	\bm{\lambda} & = (2\,\br_3 - \br_1 - \br_2) / \sqrt{6}
\end{align}
are the remaining Jacobi coordinates and
\begin{align}
	\Omega(\bm{\sigma}) & = \pi^{-\mD/4} \exp(-\sigma^2/2),
	&
	E_\Omega & = \mD/2
\end{align}
are the wave function and energy of a harmonic oscillator.

If we express the rest of the Hamiltonian in terms of the interelectronic distances instead of $\rho$, $\lambda$ and $\bm{\rho}\cdot\bm{\lambda}$, then
separation of variables reveals that the wave function $\psi$ in \eqref{eq:psi} is a solution of the Schr\"odinger-like equation
\begin{equation} \label{eq:schrodingerlike}
	\left( \nabla^2 - 2\,\kappa\,U + 2\,\epsilon - V \right) \psi = 0,
\end{equation}
where $\epsilon = E - E_\Omega$, the kinetic operator is
\begin{equation}
\begin{split}
	\frac{\nabla^2}{2}	& = \frac{\p^2}{\p r_{12}^2} + \frac{\mD-1}{r_{12}} \frac{\p}{\p r_{12}} + \frac{r_{12}^2+r_{13}^2-r_{23}^2}{2r_{12}r_{13}} \frac{\p^2}{\p r_{12} \p r_{13}}		\\
					& + \frac{\p^2}{\p r_{13}^2} + \frac{\mD-1}{r_{13}} \frac{\p}{\p r_{13}} + \frac{r_{12}^2+r_{23}^2-r_{13}^2}{2r_{12}r_{23}} \frac{\p^2}{\p r_{12} \p r_{23}}		\\
					& + \frac{\p^2}{\p r_{23}^2} + \frac{\mD-1}{r_{23}} \frac{\p}{\p r_{23}} + \frac{r_{13}^2+r_{23}^2-r_{12}^2}{2r_{13}r_{23}} \frac{\p^2}{\p r_{13} \p r_{23}},
\end{split}
\end{equation}
and the internal and external potentials are
\begin{subequations}
\begin{gather}
	U = r_{12}^{-1} + r_{13}^{-1} + r_{23}^{-1},	\\
	V = r_{12}^2 + r_{13}^2 + r_{23}^2.
\end{gather}
\end{subequations}
We also define the symmetric polynomials
\begin{subequations} \label{eq:sympol}
\begin{gather}
	\label{eq:s1}
	s_{1} = r_{12} + r_{13} + r_{23}, 
	\\
	\label{eq:s2}
	s_{2} = r_{12}\,r_{13} +r_{12}\,r_{23} + r_{13}\,r_{23},
	\\
	\label{eq:s3}
	s_{3} = r_{12}\,r_{13}\,r_{23}, 
\end{gather}
\end{subequations}
as well as the usual ``hyperradius'' 
\begin{equation}
	R = \sqrt{\frac{r_{12}^2 + r_{13}^2 + r_{23}^2}{3}},
\end{equation}
and 
\begin{equation}
	\Delta = \sqrt{s_{1}(s_{1}-2r_{12})(s_{1}-2r_{13})(s_{1}-2r_{23})},
\end{equation}
which is proportional to the area of the triangle defined by the three interelectronic distances.
These quantities will be helpful for the remainder of this Communication.

We are interested in the behavior of $\psi$ when the $r_{ij}$ are all small and, because the Laplacian is $O(r_{ij}^{-2})$, we can treat $\kappa$ as a perturbation parameter.  Expanding $\psi$ in ascending powers of $r_{ij}$ yields
\begin{equation} \label{eq:psi-expansion}
	\psi = \psi^{(0)} + \kappa\,\psi^{(1)} + \kappa^2\,\psi^{(2)} + \ldots,
\end{equation}
where the zeroth-, first- and second-order wave functions satisfy
\begin{subequations}
\begin{align}
	\label{eq:psi0}
	& \nabla^2 \psi^{(0)} = 0,
	\\
	\label{eq:psi1}
	& \nabla^2 \psi^{(1)} =  2\,U\,\psi^{(0)},
	\\
	\label{eq:psi2}
	& \nabla^2 \psi^{(2)} = 2\,U\,\psi^{(1)} - 2\,\epsilon\,\psi^{(0)}.
\end{align}
\end{subequations}
The external potential $V$ does not contribute up to second order in $r_{ij}$ (or fourth-order perturbation theory).
%----------------------------------------------------------------
\section{Results}
%----------------------------------------------------------------

%----------------------------------------------------------------
\subsection{Doublet states}
%----------------------------------------------------------------
Following Pauncz \cite{PaunczBook} and Matsen, \cite{Matsen68} the wave function of an $S = 1/2$ state is given by
\begin{equation}
\begin{split}
	^2\Phi =	& \frac{1}{\sqrt{3}} \Big[ \alpha(1) \alpha(2) \beta(3) \,^2\Psi(\bm{r}_1,\bm{r}_2|\bm{r}_3)		\\
				& \quad -  \alpha(1) \beta(2) \alpha(3) \,^2\Psi(\bm{r}_1,\bm{r}_3|\bm{r}_2)						\\
				& \quad-  \beta(1) \alpha(2) \alpha(3) \,^2\Psi(\bm{r}_3,\bm{r}_2|\bm{r}_1) \Big],
\end{split}
\end{equation}
where the vertical bar separates the spin-up and spin-down electrons, and the spatial wavefunction satisfies
\begin{subequations}
\begin{align}
	{}^2\Psi(\bm{r}_1,\bm{r}_2|\bm{r}_3) 	& = - {}^2\Psi(\bm{r}_2,\bm{r}_1|\bm{r}_3),
	\label{eq:doublet-1}
	\\
	{}^2\Psi(\bm{r}_1,\bm{r}_2|\bm{r}_3) 	& = {}^2\Psi(\bm{r}_1,\bm{r}_3|\bm{r}_2) + {}^2\Psi(\bm{r}_3,\bm{r}_2|\bm{r}_1).
	\label{eq:doublet-2}
\end{align}
\end{subequations}
Equation \eqref{eq:doublet-1} ensures that the Pauli principle is satisfied and \eqref{eq:doublet-2} (which is found in Appendix C of Ref.~\onlinecite{White70}) ensures that there is no quartet contamination.

The zeroth-order wave function is the lowest solution of Eq.~\eqref{eq:psi0} which satisfies \eqref{eq:doublet-1} and \eqref{eq:doublet-2} and,
using the Frobenius method, \cite{Frobenius, Arfken} one finds that
\begin{equation}
	^2\psi^{(0)} = r_{13}^2 - r_{23}^2.
\end{equation}
In the same way, one finds the first-order wave function
\begin{equation}
	^2\psi^{(1)} = \frac{s_{1}}{\mD^2-1} (r_{13} - r_{23}) (\mD r_{13} + \mD r_{23} - r_{12}).
\end{equation}
Solving Eq.~\eqref{eq:psi2} is difficult but it can be shown that
\begin{equation} \label{eq:psi2-doublet}
	^2\psi^{(2)} = {^2N}^{(2)} \ln(3R^2) \ {^2\chi}^{(2)} + O(R^4),
\end{equation}
where 
\begin{equation} \label{eq:chi-doublet}
	^2\chi^{(2)} = (2r_{12}^2 - r_{13}^2 - r_{23}^2 ) \ {^2\psi}^{(0)}.
\end{equation}
These zeroth-, first- and second-order wave functions agree with White and Stillinger's hyperspherical 3D results. \cite{White71}

The general second-order coefficient
\begin{equation}
\label{eq:N2}
	N^{(2)} = \int 2\,U\,\psi^{(1)}\,\chi^{(2)} d\bm{n}
\end{equation}
can be found by integrating over the hypersphere 
\begin{equation}
	\bm{n} = \left( \frac{\rho^2 - \lambda^2}{R^2}, \frac{2 \bm{\rho} \cdot \bm{\lambda}}{R^2},  \frac{2 \left| \bm{\rho} \times \bm{\lambda} \right|}{R^2} \right)
\end{equation}
of unit radius. \cite{Dmitrasinovic14}
In the 2D case, we find 
\begin{align}
	^2N^{(2)}	& = \frac{3\pi}{8} \left[3\K(\tfrac{8}{9}) - 3\E(\tfrac{8}{9}) - 114\K(\tfrac{1}{4}) + 130\E(\tfrac{1}{4}) \right]				\notag	\\
				& \approx 3.344854,
\end{align}
where $\text{K}(x)$ and $\E(x)$ are the complete elliptic integrals of the first and second kind. \cite{NISTbook}
In the 3D case, we find
\begin{equation} \label{eq:2N3D}
	^2N^{(2)} = \frac{27 \pi^2}{40} ( 11 \sqrt{3} - 6 \pi ) \approx 1.352401,
\end{equation}
which disagrees with White and Stillinger's value. \footnote{Although our normalization differs slightly from that of White and Stillinger, we could not reproduce the value given by Eq.~(60) of Ref.~\onlinecite{White71}.}

%----------------------------------------------------------------
\subsection{Quartet states with $M_S = 1/2$}
%----------------------------------------------------------------
The 3PC behavior of quartet state wave functions has not been studied before.  The wave function of an $S = 3/2$, $M_S = 1/2$ state is
\begin{align}
	^4\Phi_{1/2} = \frac{1}{\sqrt{3}} \Big[ \alpha(1) \alpha(2) \beta(3) + \alpha(1) \beta(2) \alpha(3)	\notag	\\
					+ \beta(1) \alpha(2) \alpha(3) \Big] \ ^4\Psi(\bm{r}_1,\bm{r}_2,\bm{r}_3)
\end{align}
where the spatial wave function is antisymmetric, i.e.
\begin{equation}
\begin{split}
	{^4\Psi}(\bm{r}_1,\bm{r}_2,\bm{r}_3) 	& = + {^4\Psi}(\bm{r}_2,\bm{r}_3,\bm{r}_1) = + {^4\Psi}(\bm{r}_3,\bm{r}_1,\bm{r}_2)
							\\
							& = -  {^4\Psi}(\bm{r}_1,\bm{r}_3,\bm{r}_2) = -  {^4\Psi}(\bm{r}_2,\bm{r}_1,\bm{r}_3) 
							\\
							& = -  {^4\Psi}(\bm{r}_3,\bm{r}_2,\bm{r}_1).
\end{split}
\end{equation}

Using the same approach as above, we seek antisymmetric zeroth- and first-order wave functions and find
\begin{equation}
	^4\psi_{1/2}^{(0)} = ( r_{12}^2 - r_{13}^2 ) ( r_{12}^2 - r_{23}^2 ) ( r_{13}^2 - r_{23}^2 ),
\end{equation}
and
\begin{align}
	^4\psi_{1/2}^{(1)}	& = \frac{( r_{12} - r_{13} ) ( r_{12} - r_{23} ) ( r_{13} - r_{23} ) }{(\mD+1)(\mD+3)(\mD+5)}	\notag	\\
						& \times (c_{020} s_2^2 + c_{101} s_1 s_3 + c_{210} s_1^2\,s_2 - c_{400} s_1^4 ),
\end{align}
where 
\begin{gather}
	c_{020} = 8/5,		\quad	\alert{c_{101} = 3/5 - (\mD+4)(\mD+6),} 	\\
	c_{210} = (\mD+3)(\mD+6)+\frac{2}{5},	\quad	c_{400} = \mD+3,
\end{gather}
and $s_1$, $s_2$ and $s_3$ are given by \eqref{eq:sympol}.
The second-order wave function is
\begin{equation}
\label{eq:psi2-quartet-12}
	^4\psi_{1/2}^{(2)} = {^4N}_{1/2}^{(2)} \ln ( 3R^2 ) \ ^4\chi_{1/2}^{(2)} + O(R^8),
\end{equation}
where
\begin{equation} \label{eq:chi-quartet-12}
	^4\chi_{1/2}^{(2)} = \frac{\mathcal{A}\,(r_{12}^2 r_{13}^2+r_{12}^2 r_{23}^2+r_{13}^2 r_{23}^2) - \mathcal{B}\,\Delta^2}{r_{12}^2+r_{13}^2+r_{23}^2}\ ^4\psi_{1/2}^{(0)},
\end{equation}
and 
\begin{align}
\label{eq:AB-12}
	\mathcal{A} & = \frac{\mD-1}{\mD+11},		
	&
	\mathcal{B} & = \frac{\mD+5}{\mD+11}.
\end{align}
The differences between \eqref{eq:chi-doublet} and \eqref{eq:chi-quartet-12} are interesting.  
In particular, the second-order wave function of the quartet, unlike that of the doublet, is dimension-dependent, and the logarithmic singularity for the quartet appears at order $R^8 \ln R$, rather than $R^4 \ln R$. 
This agrees with the prediction by White and Stillinger \cite{White71} that the non-analytic terms for the quartet state are of higher order.

In the 2D case, we use \eqref{eq:N2} to find
\begin{align}
\begin{split}
	^4N_{1/2}^{(2)}	& = \frac{243\pi}{6522880} \left[48171\K(\tfrac{1}{4}) - 54572\E(\tfrac{1}{4})\right] 		\\
				& \approx 0.131306.
\end{split}
\end{align}
In the 3D case, we find 
\begin{equation}
	^4N_{1/2}^{(2)} =  27 \pi^2 \left( \frac{11 \pi}{1280} - \frac{7641\sqrt{3}}{501760} \right) \approx 0.165672.
\end{equation}

%----------------------------------------------------------------
\subsection{Quartet states with $M_S = 3/2$}
%----------------------------------------------------------------
The wave function of an $S = 3/2$, $M_S = 3/2$ state is given by \cite{Herrick75b, Nodes15}
\begin{equation}
	^4\Phi_{3/2} = \alpha(1) \alpha(2) \alpha(3) D(\bm{r}_1,\bm{r}_2,\bm{r}_3)\,^4\Psi(\bm{r}_1,\bm{r}_2,\bm{r}_3),
\end{equation}
where
\begin{equation}
	D(\bm{r}_1,\bm{r}_2,\bm{r}_3) =	\begin{cases}
									\begin{vmatrix}
										x_1		&	y_1		&	1		\\
										x_2		&	y_2		&	1		\\
										x_3		&	y_3		&	1		\\
									\end{vmatrix}						&	\text{for $\mD = 2$},	\\
									\begin{vmatrix}
										x_1		&	y_1		&	z_1		\\
										x_2		&	y_2		&	z_2		\\
										x_3		&	y_3		&	z_3		\\
									\end{vmatrix}						&	\text{for $\mD = 3$},
								\end{cases}
\end{equation}
and $^4\Psi$ is a symmetric solution of \eqref{eq:schrodingerlike} in dimension $\mD+2$.
This ``interdimensional degeneracy'' has been used many times in the past, especially to calculate the energy of excited states in atomic systems. \cite{Doren86}
See Refs.~\onlinecite{Herrick75a, Herrick75b, ExSpherium10, Nodes15} for more details.

Using the Frobenius method as before, one finds that the zeroth- and first-order wave functions are
\begin{align}
	^4\psi_{3/2}^{(0)} & = 1,	
	&
	^4\psi_{3/2}^{(1)} & = \frac{s_1}{\mD+1}.
\end{align}
The second-order wave function has the form
\begin{equation}
	^4\psi_{3/2}^{(2)} = {^4N}_{3/2}^{(2)} \ln ( 3R^2 ) \ ^4\chi_{3/2}^{(2)} + O(R^2),
\end{equation}
where 
\begin{equation} \label{eq:chi-quartet-32}
	^4\chi_{3/2}^{(2)} = \frac{\mathcal{A}\,(r_{12}^2 r_{13}^2 + r_{12}^2 r_{23}^2 + r_{13}^2 r_{23}^2) - \mathcal{B}\,\Delta^2}{r_{12}^2+r_{13}^2+r_{23}^2}\ ^4\psi_{3/2}^{(0)},
\end{equation}
and 
\begin{align} \label{eq:AB-32}
	\mathcal{A} & = \frac{\mD+1}{\mD+4},		
	&
	\mathcal{B} & = \frac{\mD+5/2}{\mD+4}.
\end{align}
Though similar, the logarithmic singularities in the quartet states depend on $M_S$ via the constants in \eqref{eq:AB-12} and \eqref{eq:AB-32}.
In the 2D case, we use \eqref{eq:N2} to find
\begin{equation}
	\alert{
	^4N_{3/2}^{(2)} = \frac{9\pi^3}{32} \left[ 7\E(\tfrac{8}{9}) - 3\K(\tfrac{8}{9}) \right] \approx 1.834021.
	}
\end{equation}
In the 3D case, we find
\begin{equation}
	\alert{
	^4N_{3/2}^{(2)} = \frac{3\pi^2}{35}\left( 15\sqrt{3} - 8\pi \right) \approx 0.717397.
	}
\end{equation}

\begin{figure*}
\includegraphics[width=0.32\textwidth]{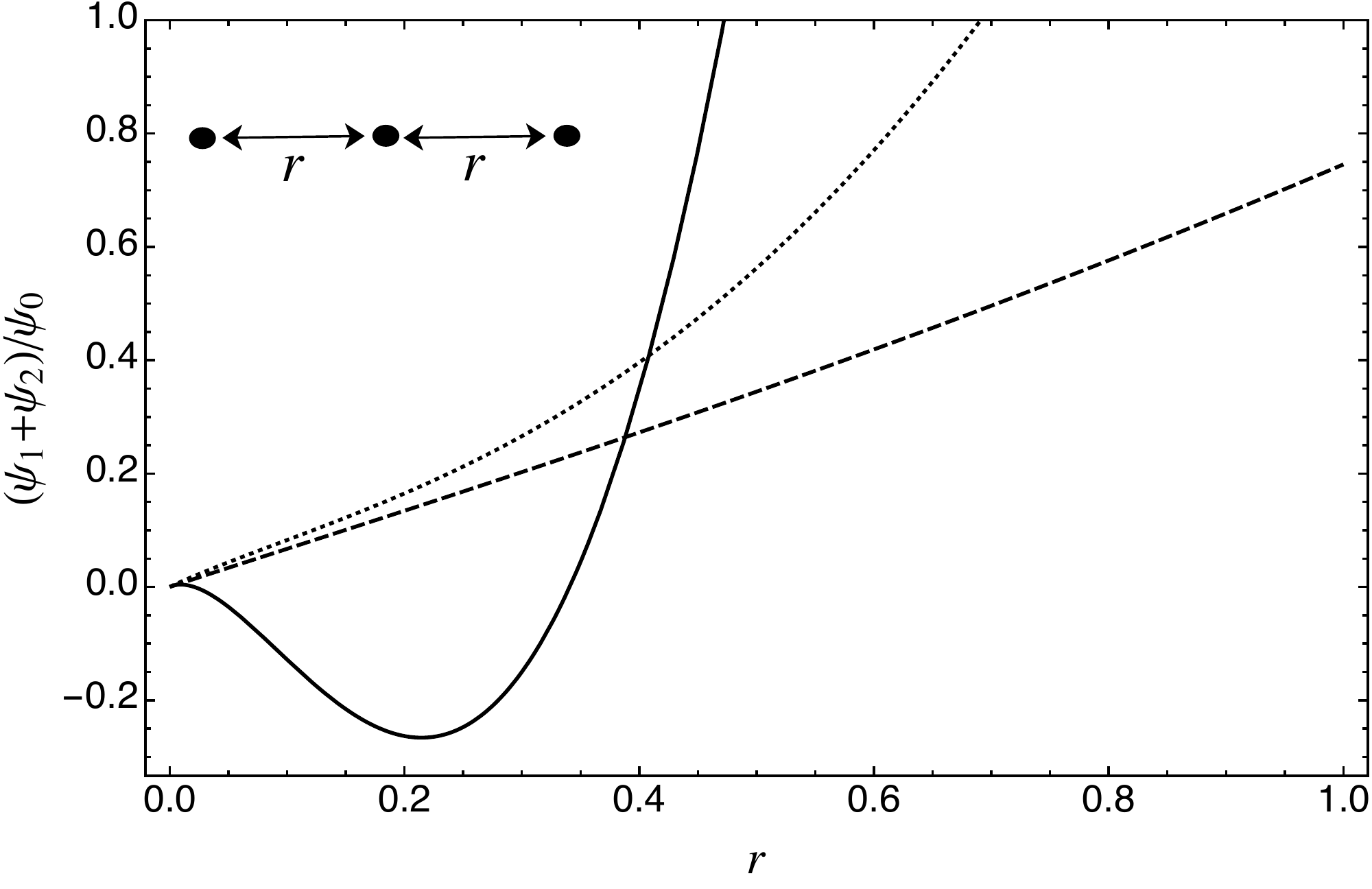}
\includegraphics[width=0.32\textwidth]{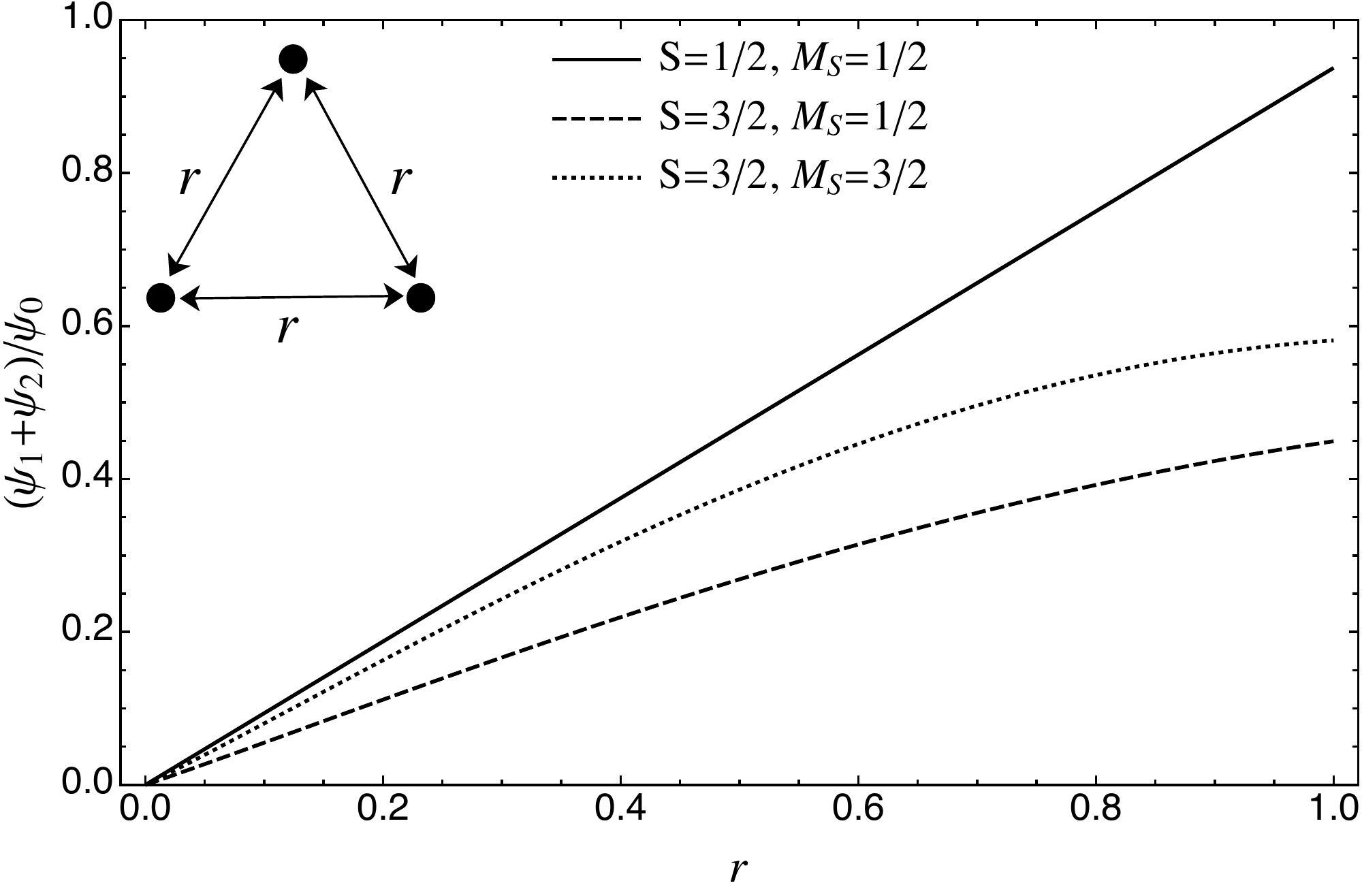}
\includegraphics[width=0.32\textwidth]{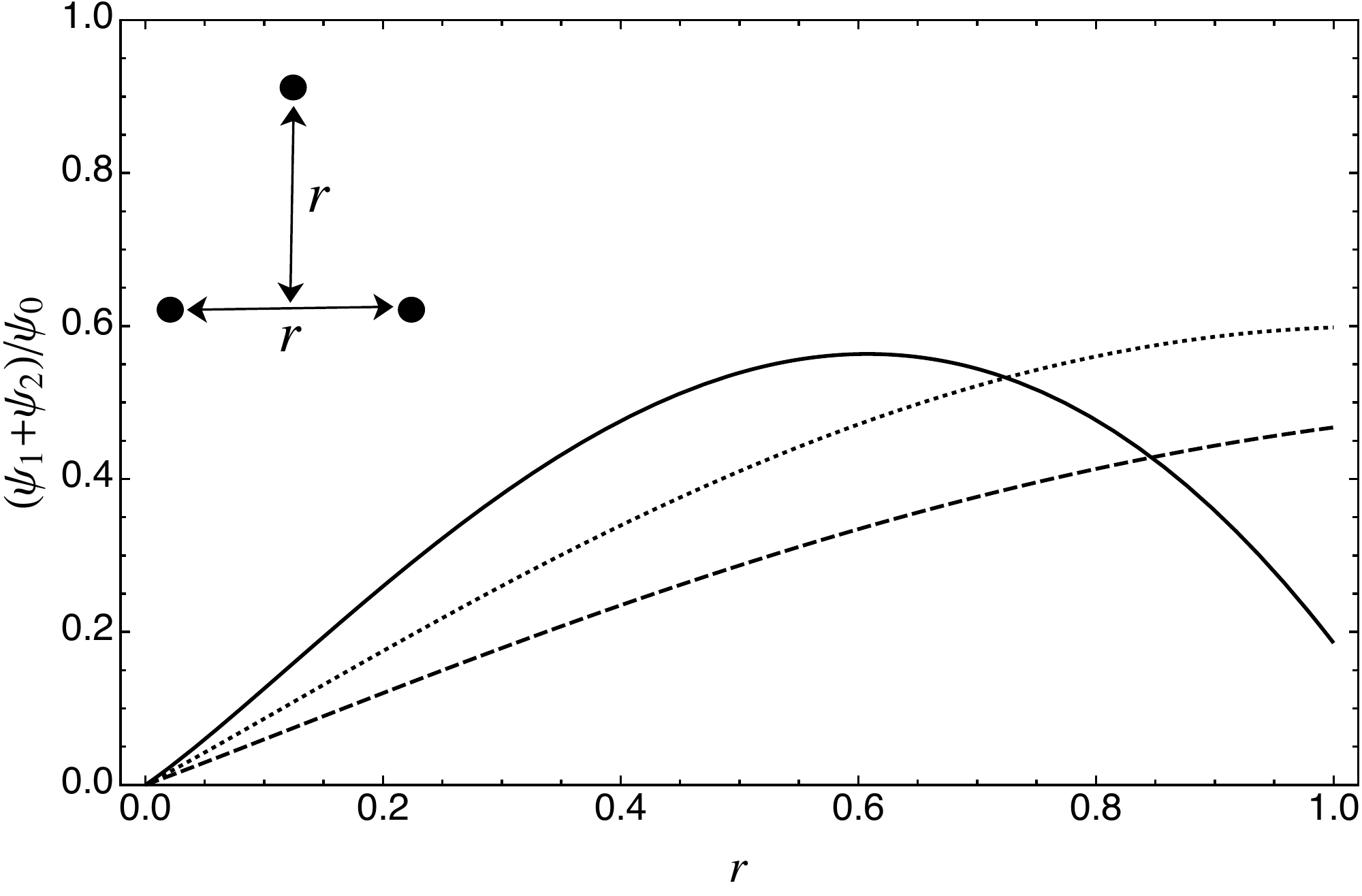}
\caption{
\label{fig:cusp}
\alert{$(\psi_1+\psi_2)/\psi_0$ at the vicinity of the 3PC for the 3D doublet and quartet states with collinear (left), equilateral (center) and isosceles (right) arrangements.
}}
\end{figure*}

\alert{In Fig.~\ref{fig:cusp}, we have represented $(\psi_1+\psi_2)/\psi_0$ for the 3D doublet and quartet states at the 3PC point for various arrangements.}

%----------------------------------------------------------------
\section{Conclusion}
%----------------------------------------------------------------
In this Communication, we have shown that the exact wave function at the three-electron coalescence point for various spin states of three electrons in a two- or three-dimensional harmonic well diverges logarithmically.
Our results should be valuable for explicitly correlated calculations and for QMC methods where the local energy discontinuity at the three-electron coalescence point could be removed by including logarithmic terms.

%----------------------------------------------------------------
\begin{acknowledgments}
%----------------------------------------------------------------
P.F.L.~thanks the Australian Research Council for a Discovery Early Career Researcher Award (Grant No.~DE130101441) and a Discovery Project grant (DP140104071).
P.M.W.G.~thanks the Australian Research Council for funding (Grants No.~DP120104740 and DP140104071). 
P.F.L.~and P.M.W.G.~also thank the NCI National Facility for generous grants of supercomputer time. 
\end{acknowledgments}

\end{document}